\documentclass[prl,amsfonts,amssymb,amsmath,showpacs,twocolumn]{revtex4}
\usepackage{epsfig}
\date{\today}
\begin{document}

\title{Comment on ``Trouble with the Lorentz law of force: Incompatibility with 
special relativity and momentum conservation''}
\author{Daniel A.\ T.\ Vanzella}\email{vanzella@ifsc.usp.br}
\affiliation{Instituto de F\'\i sica de S\~ao Carlos,
Universidade de S\~ao Paulo, Caixa Postal 369, CEP 15980-900, 
S\~ao Carlos, Brazil}

\begin{abstract}
It has been recently argued that the Lorentz force 
is incompatible with Special Relativity and should be
amended in the presence of magnetization and polarization in order 
to avoid a paradox involving a magnet in the presence of an 
electric field. Here we stress the well-known fact among relativists
that such an incompatibility
is simply impossible and show that the appearance of such a ``paradox'' 
is a mere consequence of not fully considering the relativistic consequences of 
the covariant form of the Lorentz force.
It should be mentioned, though, that this criticism
does not invalidate the debate on which is the law of force followed
by Nature, which is an experimental issue.
\end{abstract}


\maketitle

Let us reproduce the situation
analyzed in Ref.~\cite{PRL}. Consider
a magnet at rest near an electric charge, 
also at rest. For simplicity, consider also that the 
magnetic moment of the magnet is orthogonal to the separation
from the magnet to the charge.
No net force is exchanged between the two: 
static electric charges only produce
electric fields (to which the magnet seems to be oblivious) and static magnets
only produce magnetic fields (to which the static electric charge seems to be oblivious). 
The charge-magnet system stands still indefinitely.

Now, consider the same system as seen by
an inertial observer moving  parallel
to the separation between the charge and the magnet. The now
moving magnet becomes polarized with an electric dipole moment orthogonal to
its magnetic moment and to its 
velocity~\footnote{We shall see this later.}.
Therefore, the electric field produced by the (moving) 
charge will exert a torque which, Ref.~\cite{PRL} argues, would try to rotate
the magnet, trying to align its electric dipole moment to the electric field.
This clearly contradicts the fact that in the rest frame of the system the
magnet stands still.

The way out of this paradoxical situation, according to Ref.~\cite{PRL}, is to 
amend the
Lorentz force in the presence of magnetization and polarization. 
Although the alternative formula proposed, known as the Einstein-Laub law, may have its value,
the alleged motivation for this replacement does not stand.
As is well known, the
Lorentz force  can be put in a covariant form.
Anyone familiar with the geometrical formulation of Special Relativity and the 
principle
of special covariance knows that a covariant law {\it cannot}
lead to incompatible descriptions of the same phenomenon
in different inertial frames. 
This is so because once a physical law is formulated in a 
specially covariant way, one can analyze the whole phenomenon
in which it plays a role as taking place on
the four-dimensional Minkowski spacetime; no need to adopt {\it any}
inertial frame.
Different observers, who perceive the spacetime split 
differently into
space and time,  give different descriptions
of this {\it one} phenomenon based on the projections
of physical quantities onto their different 
time and space directions. But obviously all these
descriptions are connected by the same four-dimensional 
view and, therefore, cannot be inconsistent with each other. Either the proposed
covariant law accounts satisfactorily for the phenomenon in {\it all} inertial frames or
in {\it none}.
Any indication on the contrary 
points rather to
a misuse of 
relativistic arguments.

\begin{figure}[htbp]
\centering
\epsfig{file=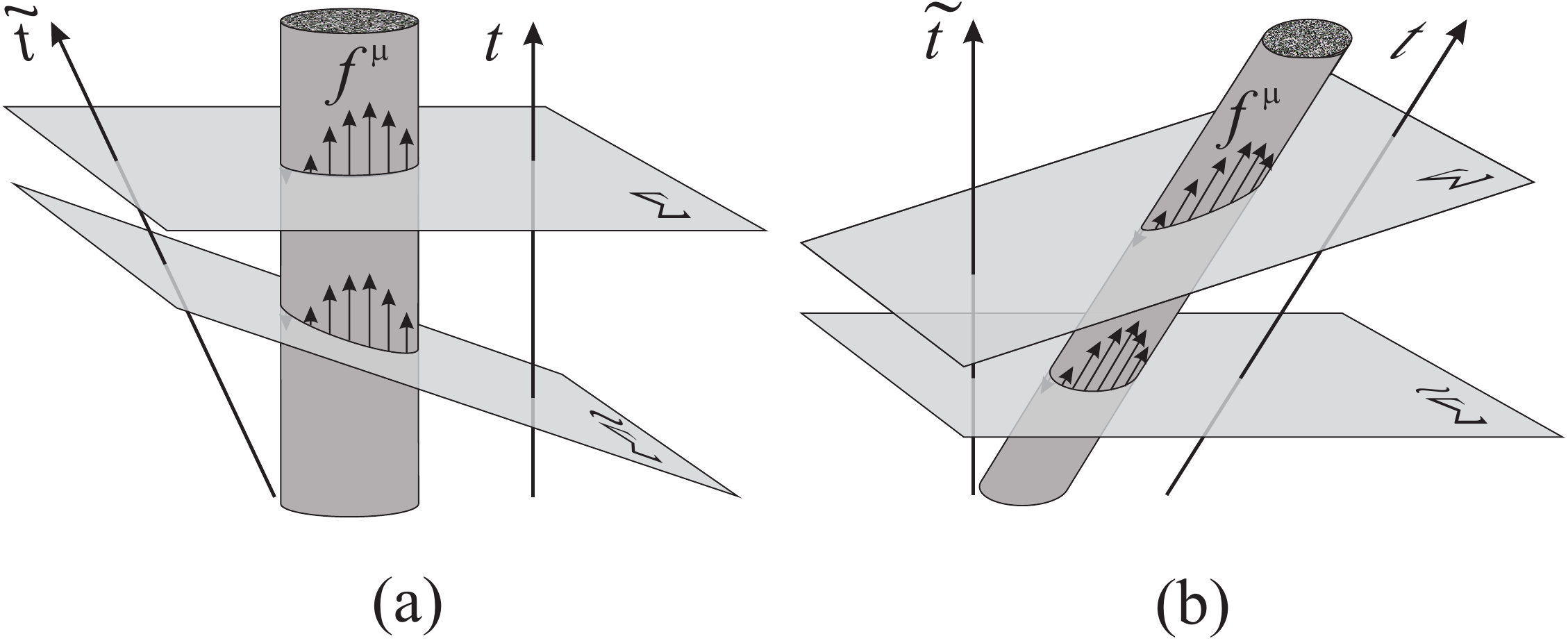, scale=0.36}
\caption{Equivalent four-dimensional representations of a cross-section of a uniformly magnetized
sphere subject to a uniform electric field: (a) Privileging the rest-frame  of the sphere, where
${t}$  is its proper-time coordinate and  ${\Sigma}$ its spatial three-surface; (b) Privileging a
``moving frame,'' where $\tilde{t}$ is its proper-time coordinate and $\widetilde{\Sigma}$ its spatial three-surface. 
The Lorentz four-force density $f^\mu= (\vec{J}\cdot \vec{E}) {u}^\mu/c^2$ 
acts on the surface of the sphere
and is future-directed where the electric field favors the surface current density (depicted in the figure) 
and past-directed otherwise (on the opposite side). It is easily seen that it has null projection on $\Sigma$ while 
having circulation (i.e., exerting torque) when projected on $\widetilde{\Sigma}$.}
\label{fig}
\end{figure}

With this in mind, let us revisit the charge-magnet system but now considering the full four-dimensional
picture enforced by Special Relativity. For simplicity, substitute the charge and its field by a
uniform electric field $\vec{E}$ and consider the magnet as being a uniformly magnetized sphere.
According to Maxwell's equations, the uniform magnetization $\vec{M}$ is caused by a net
(bound)
current density $\vec{J}$ concentrated on the surface of the sphere, circulating around
the axis of magnetization. Let  $j^\mu$ be the associated four-current density. The fact that the
magnet is  everywhere neutral in its rest frame  is encoded in $j^\mu$ being purely spatial 
according to a family ${\cal O}$ of
observers at rest with the magnet: $j^\mu {u}_\mu=0$, where ${u}^\mu$ is the four-velocity
of these observers. The uniform electric field, by its turn, is encoded in the electromagnetic tensor
$F^{\mu \nu}= E ({u}^\mu {e}_{\rm x}^\nu-{u}^\nu{e}_{\rm x}^\mu)/c$, 
where $E:=\|\vec{E}\|$ and 
${e}_{\rm x}^\mu$ gives the direction of $\vec{E}$ (${e}_{\rm x}^\mu{u}_\mu=0$, 
${e}_{\rm x}^\mu( {e}_{\rm x})_\mu=1$). Thus, the Lorentz four-force density, whose
covariant form
is given by $f^\mu = F^{\mu \nu} j_\nu/c$, 
evaluates to $f^\mu=E ({e}_{\rm x}^\nu j_\nu) {u}^\mu/c^2=(\vec{J}\cdot \vec{E})
{u}^\mu/c^2$. This means that in the rest frame of the magnet the Lorentz force has no spatial component
($f^\mu \propto {u}^\mu$; see FIG.~\ref{fig}): the Lorentz three-force is zero everywhere, 
as we knew it should be. 
However,  jumping to the conclusion that the magnet is really oblivious to the electric field in its rest frame is 
simply a mistake and stands at the root of the ``charge-magnet paradox''. 

In order to account for the effect of the 
purely time-directed Lorentz force  on the magnet, Special Relativity teaches us to look at
$\partial_\mu T^{\mu \nu}=f^\nu$, where $T_{\mu \nu}$ is the energy-momentum tensor of the magnet, which includes
matter and fields except for those responsible for $f^\mu$ (in this case, only the uniform electric field -- and its source 
-- is not included in $T_{\mu \nu}$). Although the exact form of $T_{\mu \nu}$ depends on details of the
inner structure of the magnet (stresses, energy flows, fields it generates), the net effect of $f^\mu$ 
can be obtained by first projecting the above equation in the ${u}^\mu$ direction,
$\partial _\mu {\pi}^\mu =  -f^\mu {u}_\mu/c^2
=(\vec{J}\cdot \vec{E})/c^2$ (${\pi}^\mu=-T^{\mu \nu}{u}_\nu/c^2$ is by {\it construction}
the four-momentum density of the magnet in its rest frame), then using
that the energy-momentum distribution is stationary in the rest frame of the magnet ($\partial_\mu\pi^\mu = 
{\rm div} \vec{\pi}$), 
and finally integrating the spatial part $\vec{\pi}$
of $\pi^\mu$. After some calculation one obtains that the total momentum of the magnet
{\it in its rest frame}
is given by $\vec{P}=\vec{m}\times \vec{E}/c^2$, where
$\vec{m} = 4 \pi R^3 \vec{M}/3$ is the magnet's magnetic
moment~\footnote{Dropping the simplifying assumptions that the external electric field $\vec{E}$ and the magnetization $\vec{M}$
are uniform, the result reads $\vec{P}=
\int_\Sigma \vec{M}\times \vec{E}\; d\Sigma/c^2$, where $d\Sigma$ is the proper-volume element
on the spatial three-surface $\Sigma$ (which is orthogonal to $u^\mu$).}. 
Therefore, one can easily anticipate that if the magnet moves along 
the electric field direction, its total angular momentum will be time-dependent, which will demand a net torque.

In fact,
let $\widetilde{\cal O}$ be another family of inertial observers, moving with respect 
to ${\cal O}$
with velocity $-\vec{V}$ opposite to the electric field.
Their four-velocity 
is given by $\widetilde{u}^\mu=\gamma (u^\mu-V e_{\rm x}^\mu)$,
with $\gamma:=1/\sqrt{1-V^2/c^2}$. 
According to $\widetilde{\cal O}$
the magnet is polarized with a charge-density distribution 
$\rho_{\widetilde{\cal O}}
=-j^\mu \widetilde{u}_\mu /c^2= \gamma V j_\mu e_{\rm x}^\mu/c^2=\gamma \vec{V}\cdot \vec{J}/c^2$,
leading to a dipole moment $\vec{d}=-\vec{m}\times \vec{V}/c^2$. Therefore,
the external electric field $\vec{E}$ (which is the same for $\widetilde{O}$)
exerts a torque on the magnet given by $\vec{\tau}=\vec{d}\times\vec{E}$.
It can be easily seen in FIG.~\ref{fig}(b) how the same four-force which produces no torque according to ${\cal O}$
(actually, does not even lead to a nonvanishing three-force) can exert a torque according to $\widetilde{\cal O}$.
One can check that $\vec{\tau} = \vec{V}\times \vec{P}$, which proves that the torque 
exerted
by the electric field is not used to rotate the magnet but rather is necessary to move its 
asymmetric  
momentum distribution induced by the very same electric field. 
No paradox here~\footnote{For equivalent rebuttals of the ``charge-magnet paradox'' with complementary discussions,
see Refs.~\cite{David,Cross,Saldanha2,McDonald}.}.

In summary, the four-dimensional picture
represented in FIG.~\ref{fig}
makes it very clear that the  Lorentz-force profile which the moving observers ${\widetilde{\cal O}}$
attribute as exerting a torque $\vec{\tau}$ 
on the magnet {\it  is the  same}  which the observers ${\cal O}$
in the rest frame of the magnet attribute as inducing the momentum $\vec{P}$. One cannot dismiss the latter
while considering the former, as Ref.~\cite{PRL} does.

A few concluding remarks are in order. 
The momentum $\vec{P}$ which exists in the rest frame of a 
system of particles and fields has been termed
``hidden'' in the literature because it does not reflect the (absence of) motion of the center of mass/energy
of the system. Its microscopic origin is highly dependent on the details
of the particular system and is not always easy to locate and describe:
it can be purely mechanical, due to matter flowing asymmetrically in different directions (but keeping its
center of mass/energy static), but it 
can also be due to stresses in the material. Notwithstanding, we have seen that ignoring the details
of $T_{\mu \nu}$ and the material did not prevent us to {\it calculate} the value of the hidden momentum
 induced by an external
field.  The author of Ref.~\cite{PRL} is mistaken when he
claims that hidden momenta in magnetized materials have to be ``manually'' included in order to 
avoid paradoxes when using the Lorentz force~\footnote{The author of Ref.~\cite{PRL}
claims that the hidden momentum of a uniformly magnetized sphere under the influence of
a uniform electric field should be null. His calculation assumes that the electromagnetic momentum density 
in a material medium is given by $\vec{E}\times \vec{H}/c^2$ instead of $\epsilon_0\vec{E}\times \vec{B}$. However, 
according to what we have shown here, $\vec{E}\times \vec{H}/c^2=\vec{E}\times 
(\vec{B}/\mu_0-\vec{M})/c^2$ combines both electromagnetic {\it and} hidden momenta. 
This separation has actually been suggested before~\cite{Saldanha}.}. 
On the contrary: 
this hidden momentum is consistently {\it enforced} by the Lorentz 
force.
All we have to do is fully explore its relativistic consequences.

\acknowledgments

I would like to thank Alberto Saa, Vicente Pleitez, and George Matsas for drawing my attention
to Ref.~\cite{PRL}, for useful 
discussions, and also for encouraging me in the writing of this comment. I also thank Dr.\ Masud Mansuripur 
for interesting discussions on electromagnetism in material media and the Einstein-Laub law.

\end{document}